\documentclass[fleqn,12pt,twoside]{article}
\usepackage{espcrc1}

\usepackage{graphicx}
\usepackage[figuresright]{rotating}

\newcommand{\AmS}{{\protect\the\textfont2
  A\kern-.1667em\lower.5ex\hbox{M}\kern-.125emS}}

\title{Hydro+Cascade, Flow, the Equation of State, Predictions and Data}

\author
{ 
  D. Teaney\address[PHYS]{ 
     Department of Physics and Astronomy, SUNY at Stonybrook, Stonybrook, NY 11794-3800
  }, J. Lauret\address[CHEM]{
     Department of Chemistry, SUNY at Stonybrook, Stonybrook, NY 11794-3800
  }, 
  E.V. Shuryak\addressmark[PHYS]
}

\begin{document}

\maketitle

\begin{abstract}
A Hydro+Cascade 
model has been used to describe radial and elliptic
flow at the SPS and successfully predicted 
the radial and elliptic flow measured by the both STAR and PHENIX collaborations. 
Furthermore, a combined description of the radial and elliptic
flow for different particle species, restricts the Equation of State(EoS) 
and points towards an EoS with a phase transition to the Quark Gluon Plasma(QGP).
\end{abstract}

\section{Introduction}

Relativistic Hydrodynamics provides a link between the
Equation of State(EoS) of the excited nuclear matter and
collective observables such as elliptic flow($v_2$-s) 
and radial flow($T_{slope}$-s).
At the SPS, 
pure hydrodynamics calculations can fit the transverse mass spectra 
for almost any EoS by choosing  the freezeout temperature $T_{f}$.
However in non-central collisions,
when $v_2$ was calculated for these same EoSs 
with the same $T_{f}$-s, 
$v_{2}$ was above the data by a factor of two\cite{Kolb}.
Bass and Dumitru\cite{Bass} removed the $T_{f}$ indeterminacy 
by injecting
the particles into a microscopic transport model at a
switching temperature 
$T_{switch}\approx T_{c}\approx165\,MeV$, and cascading the
particles until they decoupled.  With the freezeout parameter
removed, the slope parameters for central PbPb collisions 
were calculated and found to agree with experimental values.
The only parameter in this approach is the total 
multiplicity in the collision.

Later, elliptic flow was calculated\cite{Teaney} in a similar
Hydro+Cascade model and $v_{2}$ was only 20\% above the data.
Since at freezeout the viscosity is 
certainly important, it was not surprising that the introduction 
of a cascade reduced
the elliptic flow.
It $was$ surprising that Hydro+Cascade could
simultaneously reproduce the elliptic $and$ radial flow for
different particle species as a function of impact parameter.
Furthermore, 
the combined analysis of radial and elliptic flow
restricted possible EoSs since 
the freezeout temperature could no longer be adjusted to make
any EoS fit any slope parameter. Roughly speaking, a soft
EoS produced too little radial flow while a hard EoS produced
too much elliptic flow. 

Now, with the EoS roughly fixed from available SPS data, parameter
free predictions were made for RHIC. 
A few of these predictions
are: (a) an increase in $v_{2}$ by approximately 40\% over the
SPS and (b) a significant increase in the radial flow.
(c) curved nucleon $m_{T}$ spectra. The preliminary
data reported in this conference have been in agreement with 
these predictions .

Ideally, the cascade should provide a kind of dual description
of the hydrodynamics. Although $T_{c}$ provides a natural
place to switch to the cascade, the results (slope parameters 
and $v_{2}$-s, lifetimes, etc.)  should be insensitive to 
the switching temperature $T_{switch}$. Unfortunately, 
Bass and Dumitru\cite{Bass} reported that the results were sensitive to the
transition surface. By incorporating chemical freezeout into 
the hydrodynamic calculation, the sensitivity to $T_{switch}$ 
was much reduced although the elliptic flow at the
SPS remained sensitive to $T_{switch}$\cite{Teaney-prog}.

If the preliminary data remain unchanged and further
predictions are verified, the hydrodynamic 
description must be taken seriously and the equilibration 
times and transport cross sections estimated from binary,
perturbative, classical parton cascades 
must be considered only a very coarse guide to the radiating, 
non-perturbative, quantum glue that makes up the initial
state.

%
\begin{figure}[tb]
\begin{minipage}{160mm}
\includegraphics[width=80mm,height=7cm]{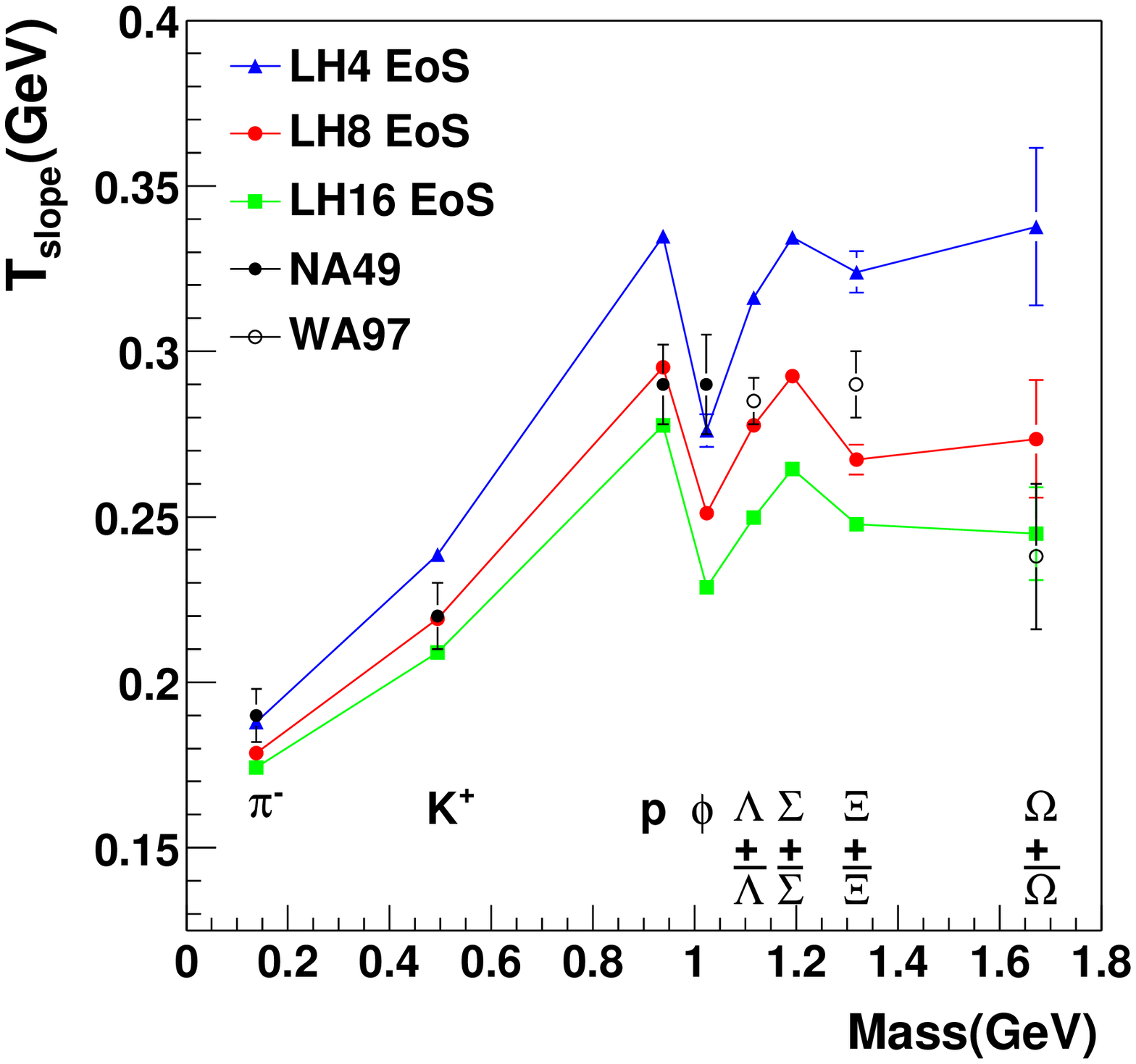} 
\hfill
\includegraphics[width=80mm,height=7cm]{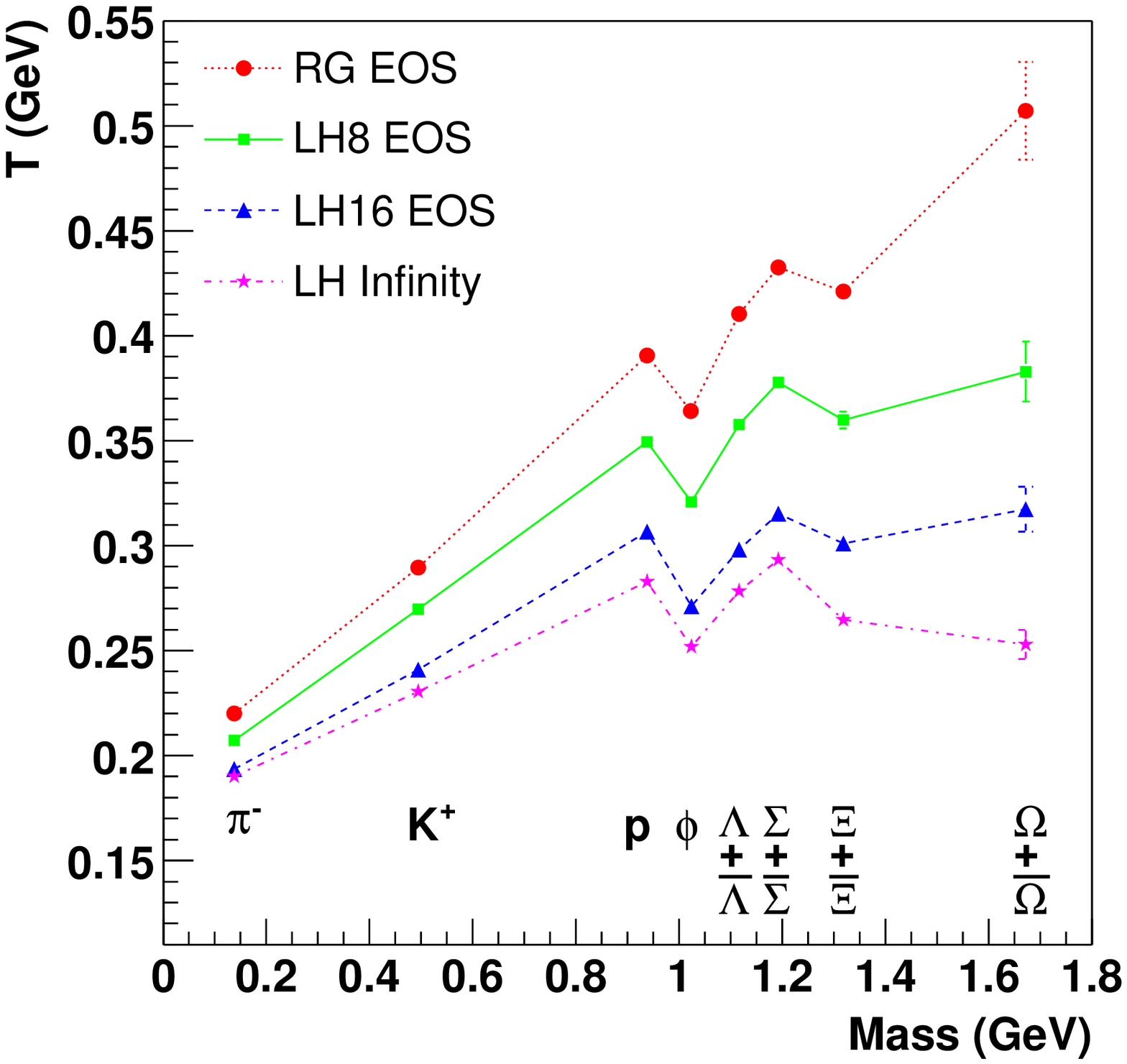}
\caption{
(a) A compilation of slope parameters(see e.g.\cite{Stachel}) 
at the SPS compared
to model predictions for different EoSs. The slope parameters
are fit from 0$ < M_{T}-m < $0.9\,GeV, corresponding to the 
WA98 acceptance.
(b) Model predictions for slope parameters at RHIC for
different EoSs. The slope parameters are fit over the range
0$ < M_{T}-m < $1.6\,GeV and do depend on the fit range used.
}
\label{EoSs}
\end{minipage}
\end{figure}
\begin{figure}[tb]
\begin{minipage}{160mm}
\includegraphics[width=80mm,height=7cm]{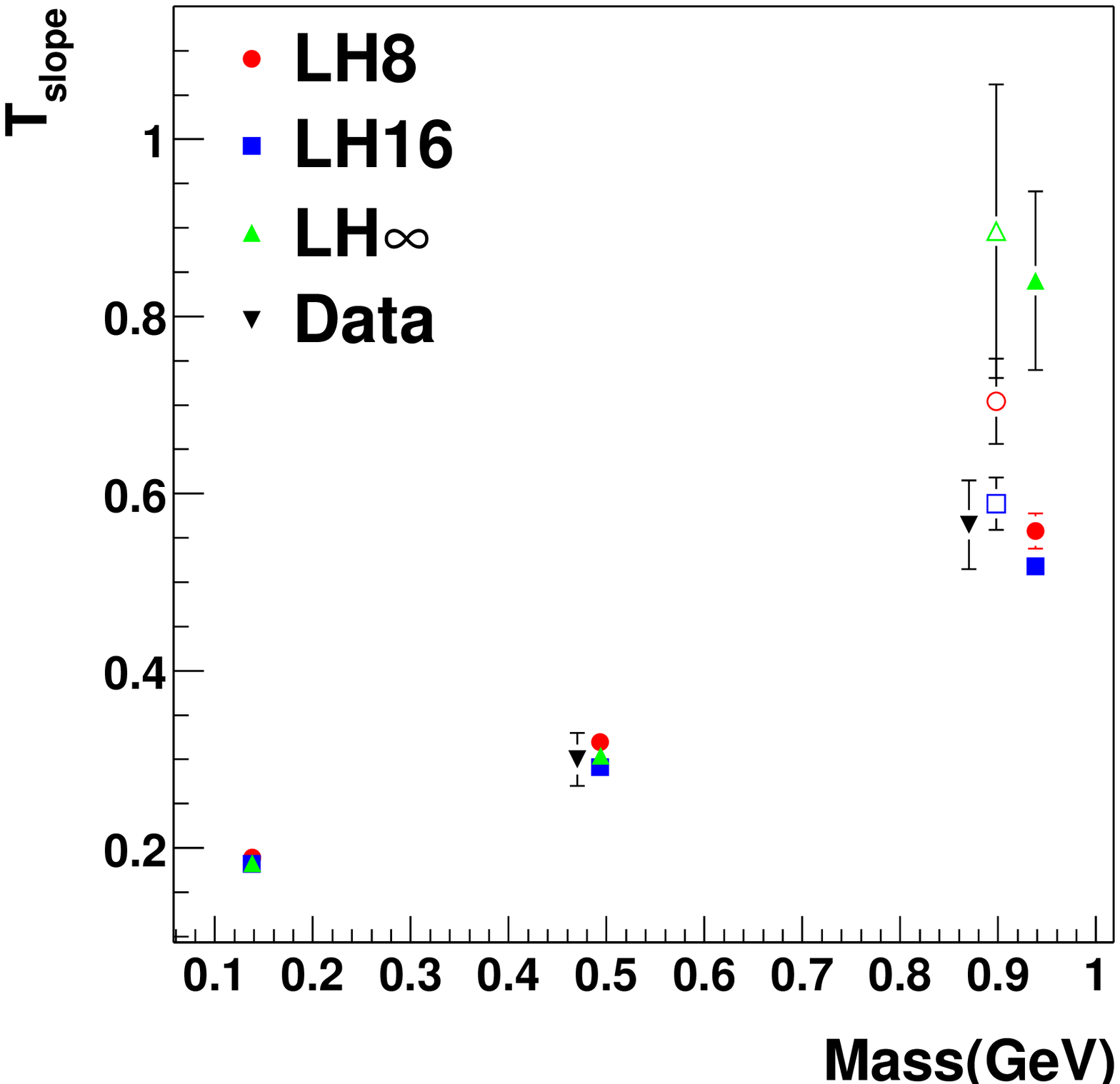} 
\hfill
\includegraphics[width=80mm,height=7cm]{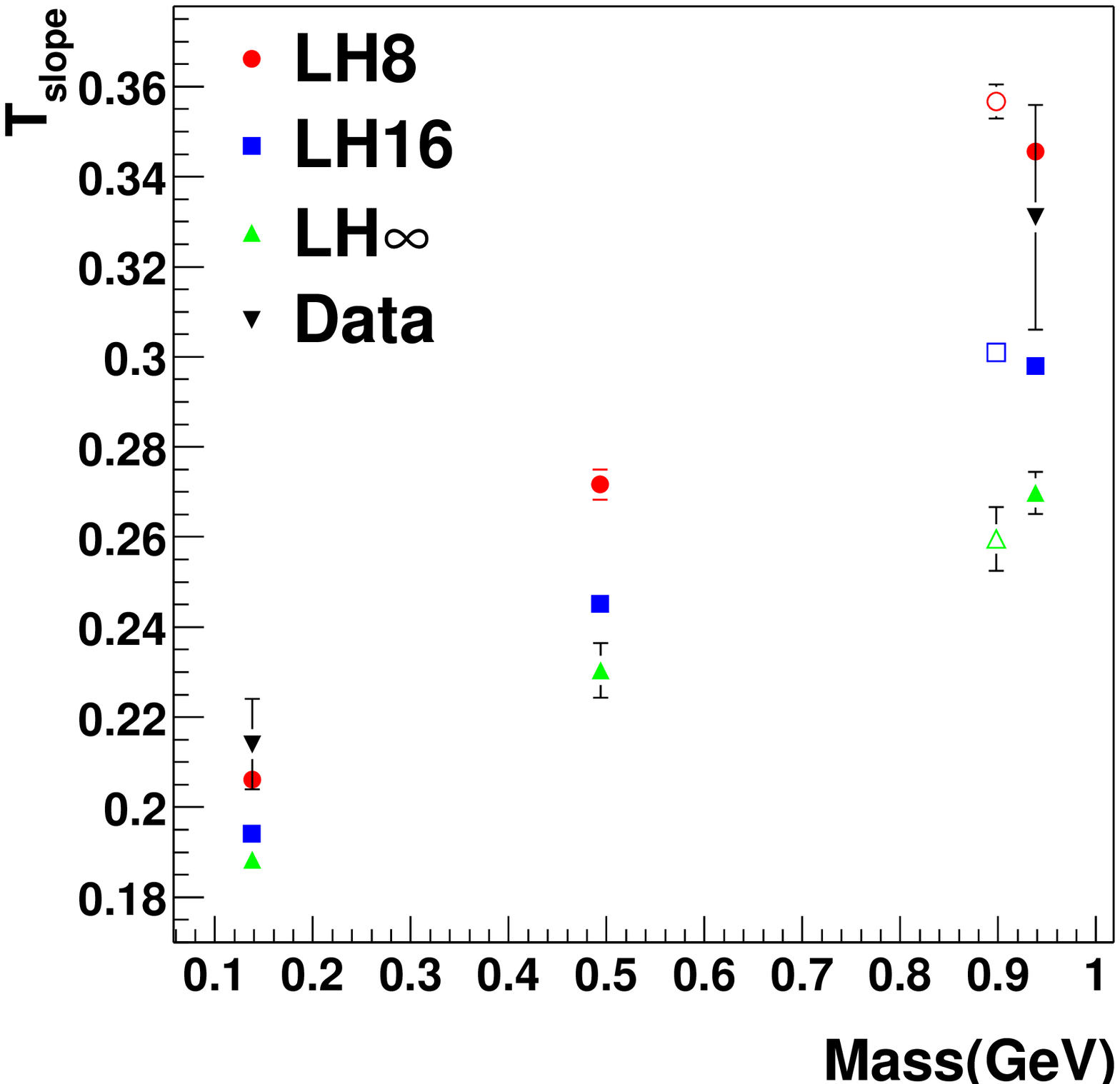}
\caption{
  Slope parameters for $\pi^{-}$, $K^{-}$, $p$ and $\bar{p}$
  reported in these proceedings by the STAR(a) and PHENIX(b)
  collaborations\cite{NuXu,PHENIX-p}. The open(closed) symbols show
  model predictions for anti-protons(protons).  
  The STAR collaboration fits the $\pi^{-}$, $K^{-}$ and $\bar{p}$  
  spectra over the ranges,  
  0.12\,GeV$ < M_{T}-m < $0.45\,GeV,  0.04\,GeV$ < M_{T}-m < $0.34\,GeV and 0.04\,GeV$ < M_{T}-m < $0.45\,GeV  
  respectively.
  The PHENIX collaboration fits the $\pi^{-}$ and $p$ spectra over
  the ranges, 0.19\,GeV$ < M_T - m < $0.87\,GeV and 0.175\,GeV$ < M_T-m < $2.2\,GeV respectively. 
}
\label{slope_dat}
\end{minipage}
\end{figure}
%

\section{The EoS, Flow, Predictions and Data} 
Below a family of EoSs with a first order 
phase transition are studied and are labeled by the Latent Heat(LH).
LH4, LH8, $\cdots$ label EoSs with a latent heats of $0.4\,GeV/fm^{3}$,
$0.8\,GeV/fm^{3}$, $\cdots$.
LH$\infty$ is studied as a limiting case. 
A Resonance Gas(RG) EoS (which does not have 
a phase transition) is also studied.

First in Fig.\,\ref{EoSs}(a), the measured slope parameters at the SPS
are compared to model predictions for different EoSs.
LH8 gives the best description of the available spectra.
LH4 is too stiff (the slope parameters are too high) and 
LH16 is too soft (the slope parameters are too low). 
The $T_{slope}$-s for a RG EoS are comparable to LH4,
and the $T_{slope}$-s for LH$\infty$ are comparable to 
LH16. 

Although LH8 gives the best fit to SPS spectra, 
the model-data discrepancy 
for the other EoSs is not large. As the collision energy is 
increased from the SPS to RHIC the slope parameters all
increase (see e.g. \cite{Bass,Teaney}). Since 
at high energies the importance of the QGP phase increases,
the differences between these EoSs are magnified during
the systems evolution. In Fig.\,\ref{EoSs}(b), the
model predictions at RHIC collision 
energies are shown for different EoSs. Note,  
the spectra are curved and the parameterization in terms
of slope parameters is only schematic. LH$\infty$, with
no QGP push, generates only small slope parameters. 
The differences between the  EoSs is clear in the flow
of the $\Omega$.

With these predictions, a comparison to the first 
RHIC data is made in 
Fig.\,\ref{slope_dat}.
The $M_{T}$ spectra are curved (see \cite{NuXu}) and therefore
the STAR and PHENIX collaborations measure quite different slope
parameters. The STAR collaboration 
fits the observed spectra in a low $M_{T}$ range  and
measures large slopes,
while the PHENIX collaboration fits  
in a high $M_{T}$ range and measures small slopes. 
The best agreement with the proton and anti-proton
slope parameters of both collaborations 
is found between LH8 and LH16 . LH$\infty$ has
too much flow at small $M_{T}$ and too little flow at high $M_{T}$ 
 and therefore fails to reproduce the curvature of the $M_{T}$ spectra seen
in the data. The slope parameters reported in this 
conference implicate a strong transverse expansion.
\begin{figure}[t]
\begin{minipage}{160mm}
\includegraphics[width=80mm,height=7cm]{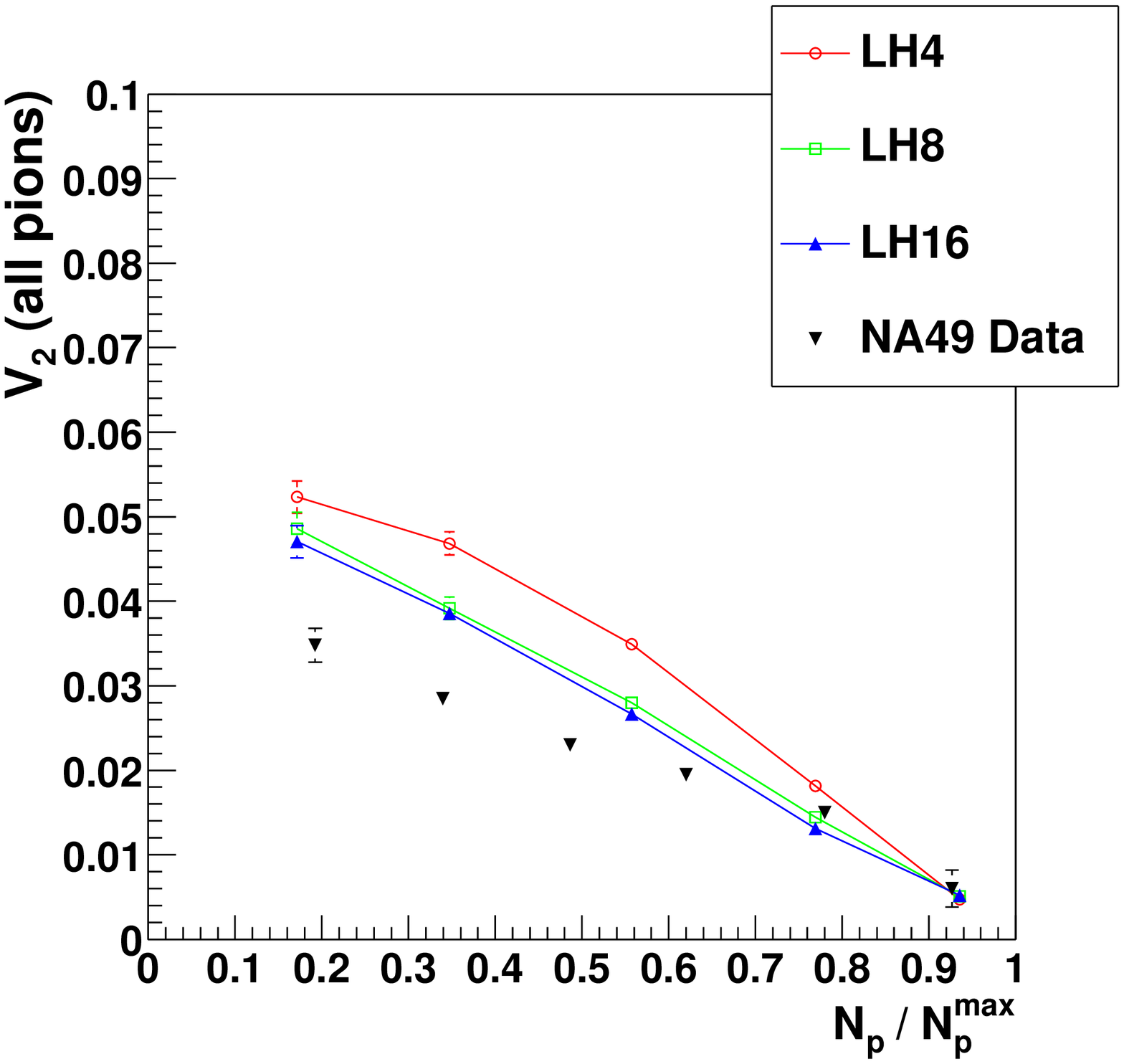}
\includegraphics[width=80mm,height=7cm]{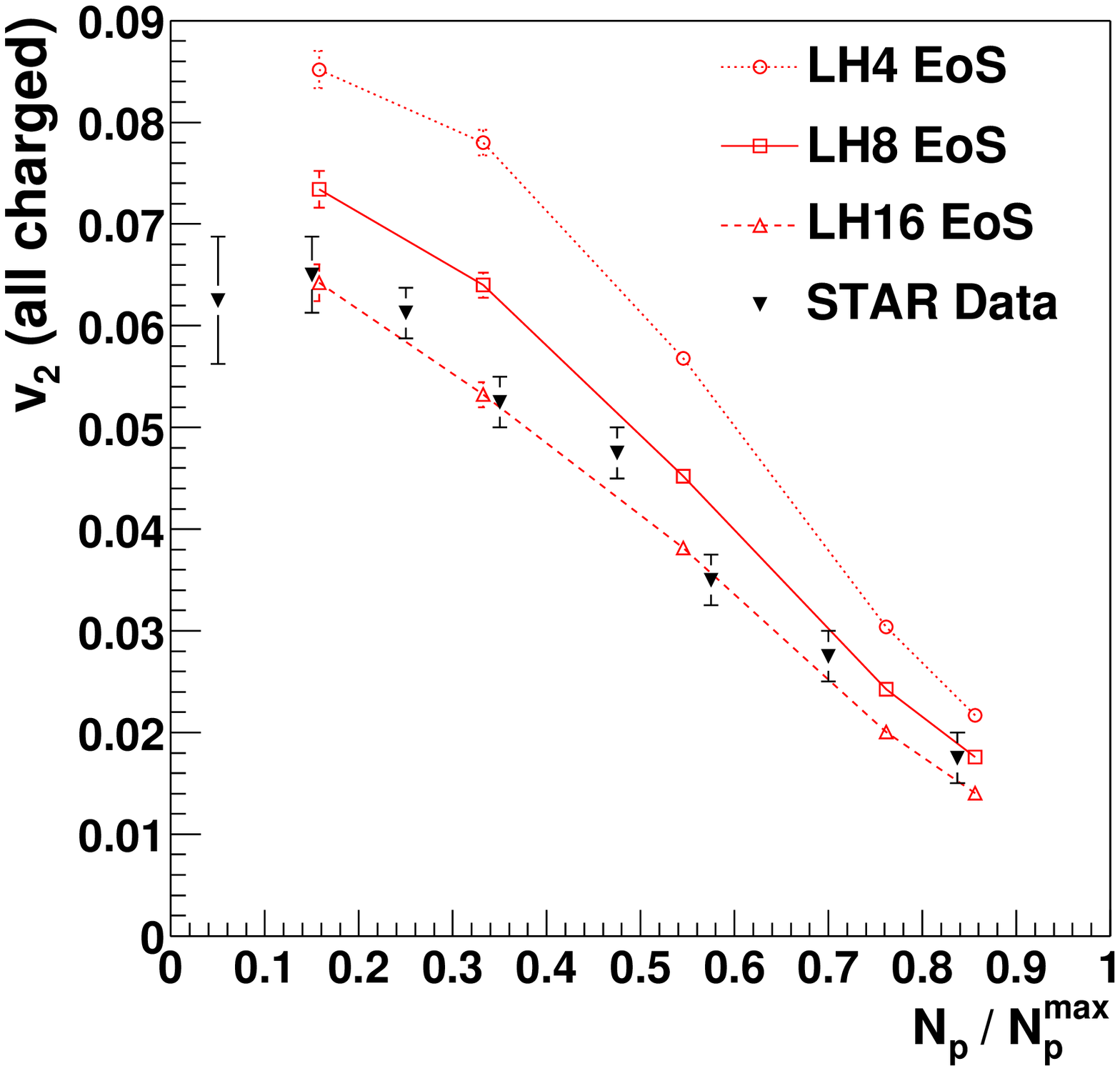} 
\vskip -1.0cm
\caption{Elliptic flow  for different EoSs as a function
of the number of participants relative to the maximum at the
SPS(a) and RHIC(b). (a)The data points are for $\pi^{-}$\cite{NA49} 
and the model points are for all pions.  
(b)The model points and data points\cite{STAR-e} are 
for all charged particles.}
\label{eflow_dat}
\end{minipage}
\end{figure}

Now, elliptic flow is studied as a function of impact parameter.
In Fig.\,\ref{eflow_dat}(a) and (b), the  elliptic  
flow of pions/charged particles at the SPS and RHIC is shown for different EoSs.
Notice that elliptic flow increases by approximately $40\%$ from
the SPS to RHIC. This prediction was borne out by the first 
STAR measurements\cite{STAR-e}. At the SPS, EoSs  
without or only a very weak phase transition (e.g. RG or LH4) produce far too much 
elliptic flow. At the SPS, LH8 and LH16 give approximately the same 
elliptic flow since the contribution to $v_{2}$ of the pure QGP phase
is small. At RHIC, the $v_{2}$-s of LH4, LH8 and LH16 begin to separate
as the QGP phase becomes increasingly significant.

{\bf Acknowledgements}.
P. Kolb is thanked for critically reading this manuscript. This work is partly supported by
US DOE grant No. DE-FG02-88ER40388 and grant No. DE-FGO2-87ER 40331. 



\begin{thebibliography}{9}
\bibitem{Kolb}
   P. Kolb, J. Sollfrank, U. Heinz, Phys. Lett. {\bf B459}, 667  (1999).
\bibitem{Bass}
   S. Bass and A. Dumitru,  Phys. Rev. {\bf C61}, 064909 (2000).
\bibitem{Teaney} 
   D. Teaney, J. Lauret, E.V. Shuryak, Phys. Rev. Lett. in press.
   nucl-th/0011058.
\bibitem{Teaney-prog} D. Teaney, J. Lauret, E.V. Shuryak, in progress.
\bibitem{NuXu} Nu Xu, for the STAR Collaboration, these proceedings.
\bibitem{STAR-e} R. Snellings, for the STAR Collaboration,
these proceedings.
\bibitem{PHENIX-p} J. Velkovska,  for the PHENIX Collaboration, these
proceedings.
\bibitem{NA49}
  NA49 Collaboration, H.~Appelsh\"{a}user {\em et al.},
  Phys. Rev. Lett.  {\bf 80}, 4136  (1998).
\bibitem{Stachel}
  see J. Stachel, nucl-exp/9903007 for the original
  references.
\end{thebibliography}
\end{document}